\documentclass[journal=jacsat,manuscript=article]{achemso}

\usepackage{chemformula} 
\usepackage[T1]{fontenc} 

\title{Gate-controlled superconducting switch in GaSe/NbSe$_2$ van der Waals heterostructure}

\author{Yifan Ding}
\affiliation{School of Physical Science and Technology, ShanghaiTech University, Shanghai 201210, China}
\alsoaffiliation{ShanghaiTech Laboratory for Topological Physics, ShanghaiTech University, Shanghai 201210, China}
\author{Chenyazhi Hu}
\affiliation{School of Physical Science and Technology, ShanghaiTech University, Shanghai 201210, China}
\alsoaffiliation{ShanghaiTech Laboratory for Topological Physics, ShanghaiTech University, Shanghai 201210, China}
\author{Wenhui Li}
\affiliation{Institute of Physics, Chinese Academy of Sciences, Beijing, 100190, China}
\author{Lan Chen}
\affiliation{Institute of Physics, Chinese Academy of Sciences, Beijing, 100190, China}
\email{lchen@iphy.ac.cn}

\author{Jiadian He}
\affiliation{School of Physical Science and Technology, ShanghaiTech University, Shanghai 201210, China}
\alsoaffiliation{ShanghaiTech Laboratory for Topological Physics, ShanghaiTech University, Shanghai 201210, China}

\author{Yiwen Zhang}
\affiliation{School of Physical Science and Technology, ShanghaiTech University, Shanghai 201210, China}
\alsoaffiliation{ShanghaiTech Laboratory for Topological Physics, ShanghaiTech University, Shanghai 201210, China}

\author{Xiaohui Zeng}
\affiliation{School of Physical Science and Technology, ShanghaiTech University, Shanghai 201210, China}
\alsoaffiliation{ShanghaiTech Laboratory for Topological Physics, ShanghaiTech University, Shanghai 201210, China}

\author{Yanjiang Wang}
\affiliation{School of Physical Science and Technology, ShanghaiTech University, Shanghai 201210, China}
\alsoaffiliation{ShanghaiTech Laboratory for Topological Physics, ShanghaiTech University, Shanghai 201210, China}

\author{Peng Dong}
\affiliation{School of Physical Science and Technology, ShanghaiTech University, Shanghai 201210, China}
\alsoaffiliation{ShanghaiTech Laboratory for Topological Physics, ShanghaiTech University, Shanghai 201210, China}

\author{Jinghui Wang}
\affiliation{School of Physical Science and Technology, ShanghaiTech University, Shanghai 201210, China}
\alsoaffiliation{ShanghaiTech Laboratory for Topological Physics, ShanghaiTech University, Shanghai 201210, China}

\author{Xiang Zhou}
\affiliation{School of Physical Science and Technology, ShanghaiTech University, Shanghai 201210, China}
\alsoaffiliation{ShanghaiTech Laboratory for Topological Physics, ShanghaiTech University, Shanghai 201210, China}

\author{Yueshen Wu}
\email{wuysh@shanghaitech.edu.cn}
\affiliation{School of Physical Science and Technology, ShanghaiTech University, Shanghai 201210, China}
\affiliation{ShanghaiTech Laboratory for Topological Physics, ShanghaiTech University, Shanghai 201210, China}

\author{Yulin Chen}
\affiliation{School of Physical Science and Technology, ShanghaiTech University, Shanghai 201210, China}
\alsoaffiliation{ShanghaiTech Laboratory for Topological Physics, ShanghaiTech University, Shanghai 201210, China}
\alsoaffiliation{Department of Physics, Clarendon Laboratory, University of Oxford, Oxford OX1 3PU, UK}

\author{Jun Li}
\email{lijun3@shanghaitech.edu.cn}
\affiliation{School of Physical Science and Technology, ShanghaiTech University, Shanghai 201210, China}
\alsoaffiliation{ShanghaiTech Laboratory for Topological Physics, ShanghaiTech University, Shanghai 201210, China}

\begin{document}

\clearpage
\begin{abstract}
The demand for low-power devices is on the rise as semiconductor engineering approaches the quantum limit and quantum computing continues to advance. Two-dimensional (2D) superconductors, thanks to their rich physical properties, hold significant promise for both fundamental physics and potential applications in superconducting integrated circuits and quantum computation. Here, we report a gate-controlled superconducting switch in GaSe/NbSe$_2$ van der Waals (vdW) heterostructure. By injecting high-energy electrons into NbSe$_2$ under an electric field, a non-equilibrium state is induced, resulting in significant modulation of the superconducting properties. Owing to the intrinsic polarization of ferroelectric GaSe, a much steeper subthreshold slope and asymmetric modulation are achieved, which is beneficial to the device performance. Based on these results, a superconducting switch is realized that can reversibly and controllably switch between the superconducting and normal state under an electric field. Our findings highlight a significant high-energy injection effect from band engineering in 2D vdW heterostructures combining superconductors and ferroelectric semiconductors, and demonstrate the potential applications for superconducting integrated circuits.

\end{abstract}


\section{INTRODUCTION}

Superconducting devices, owing to their combination of ultralow power consumption and ultrahigh speed, hold great promise as fundamental components for quantum computing architectures and associated cryogenic control electronics\cite{nature2003, np2012, np2023, nature2023}.
Considering the limited space of the cryogenic system and the requirement for minimal interference between devices, the recently rapidly-developed experimental techniques, such as the magnetic fields\cite{magnetic1, magnetic2}, light pulses\cite{light1, light2}, microwave \cite{JJ1, JJ2}, or plasmon excitations\cite{plasma}, are all more challenging to implement than electric fields. In this context, it is of great significance to develop a superconducting `triode' device that can be electrically tuned between a superconducting state and a normal resistive state through a gate electrode\cite{PRA2020, ne2019, Phonon3, PRA2024}.

\begin{figure*}[!htbp]
\centering
\includegraphics[width=1 \linewidth]{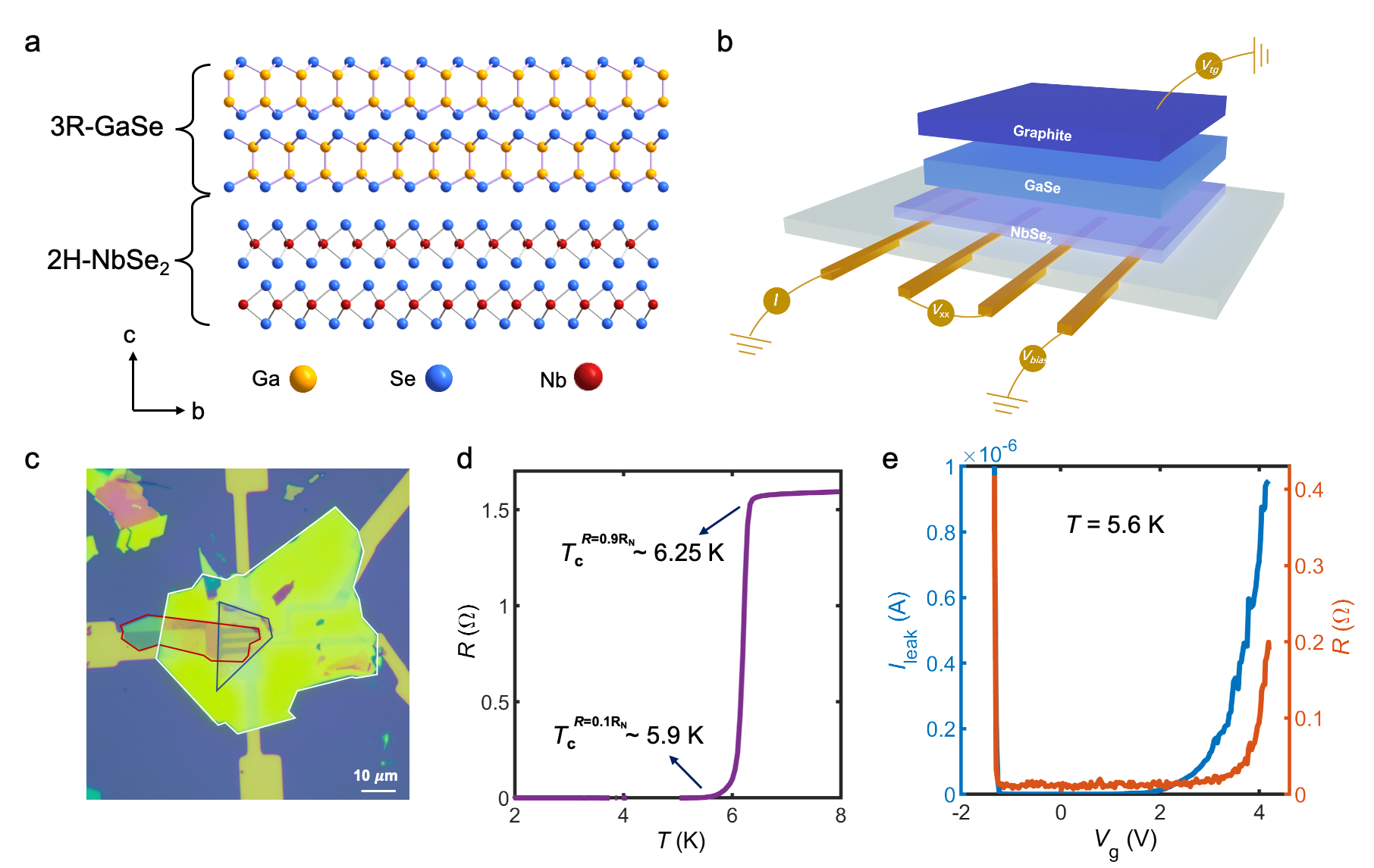}
\caption{{\label{fig1} \textbf{Few-layer NbSe$_2$ gated with a GaSe flake.} \textbf{(a)} Cross-sectional view of the GaSe/NbSe$_2$ heterostructure. The diagram employs yellow, blue, and red spheres to represent Ga, Se, and Nb atoms, respectively. \textbf{(b)} A schematic image of the stacking order of thin graphite flake, GaSe flake, and few-layer NbSe$_2$. The Ti/Au electrodes are embedded in the SiO$_2$/Si substrates, ensuring effective contact with the few-layer NbSe$_2$. \textbf{(c)} An optical image of the device. The blue, red and white solid lines represent few-layer NbSe$_2$, graphite flake, and GaSe flake, respectively. The scale bar corresponds to 10 $\mu$m. \textbf{(d)} The $R$-$T$ curve of the few-layer NbSe$_2$. \textbf{(e)} The $I_{\text{leak}}$-$V_{\text{g}}$ curve (in blue) and $R$-$V_{\text{g}}$ curve (in orange) of the GaSe/NbSe$_2$ heterostructure at a temperature of 5.6 K. As the gate voltage increases, NbSe$_2$ exhibits a transition from the superconducting state to the normal state.}}
\label{fig:fig1}
\end{figure*}

Among the various types of superconductors, 2H-NbSe$_2$ has been recognized as a well-known 2D superconductor, capable of maintaining superconductivity down to sub-nanometer thickness (monolayer limit)\cite{Ising, CVD}. Notably, it exhibits a range of unique physical properties, such as the superconductivity with Ising spin-orbit coupling protection\cite{Ising, FFLO}, quantum metallic state (Bose metal)\cite{metal}, and competition with charge density wave order\cite{CDW}. More importantly, thanks to the thermodynamic stability of 2H phase (Fig. \ref{fig1}(a)), 2H-NbSe$_2$ can be fabricated into a range of micro/nano functional devices\cite{NSring, NSwire, NSbridge}. Nevertheless, owing to the common properties of most superconductors with high charge carrier density, the modulation of superconducting properties by direct electric field effects have proven to be quite challenging in 2H-NbSe$_2$\cite{gating, gating2, NS-STO}. Hence, there is a pressing requirement to explore a simpler and more efficient approach for modulating the functional devices based on 2H-NbSe$_2$. 


The injecting of high-energy electrons into a certain material via an electric field, thereby manipulating its superconducting properties, has recently been experimentally verified in superconducting thin films\cite{Emission1, Phonon1, Phase2} and nanowires\cite{Emission2, Emission3, Emission4, Phonon2, Phonon3, Phase1}. In order to realize this effect, a suitable energy barrier is required. Recently, $\gamma$-GaSe, a III-VI post-transition metal chalcogenides compound semiconductor with a band gap of approximately 2 eV (Fig. \ref{fig1}(a)), has garnered significant attention due to its spontaneous ferroelectric polarization\cite{GaSe}. Therefore, employing the GaSe flake as a barrier to construct a gate-controlled superconducting device appears promising, and the role of ferroelectric polarization on modulating physical properties of NbSe$_2$ is particularly intriguing.



In this work, we studied on a gate-controlled superconducting manipulation in few-layer NbSe$_2$ through fabricating a NbSe$_2$/GaSe vdW device, as shown in Fig. \ref{fig1}(b) and Fig. \ref{fig1}(c). 
An asymmetric $R$-$V_{\text{g}}$ curve in few-layer NbSe$_2$ was observed due to the spontaneous ferroelectric polarization of GaSe flake. By increasing the injection of high-energy electrons, we were able to achieve a modulation range of 0.4 K for the superconducting critical temperature of NbSe$_2$. Moreover, we implemented a gate-controlled superconducting switch in this device, showcasing its potential for applications in superconducting circuits.

\section{RESULTS AND DISCUSSIONS}
\begin{figure*}[!htbp]
\centering
\includegraphics[width=1\linewidth]{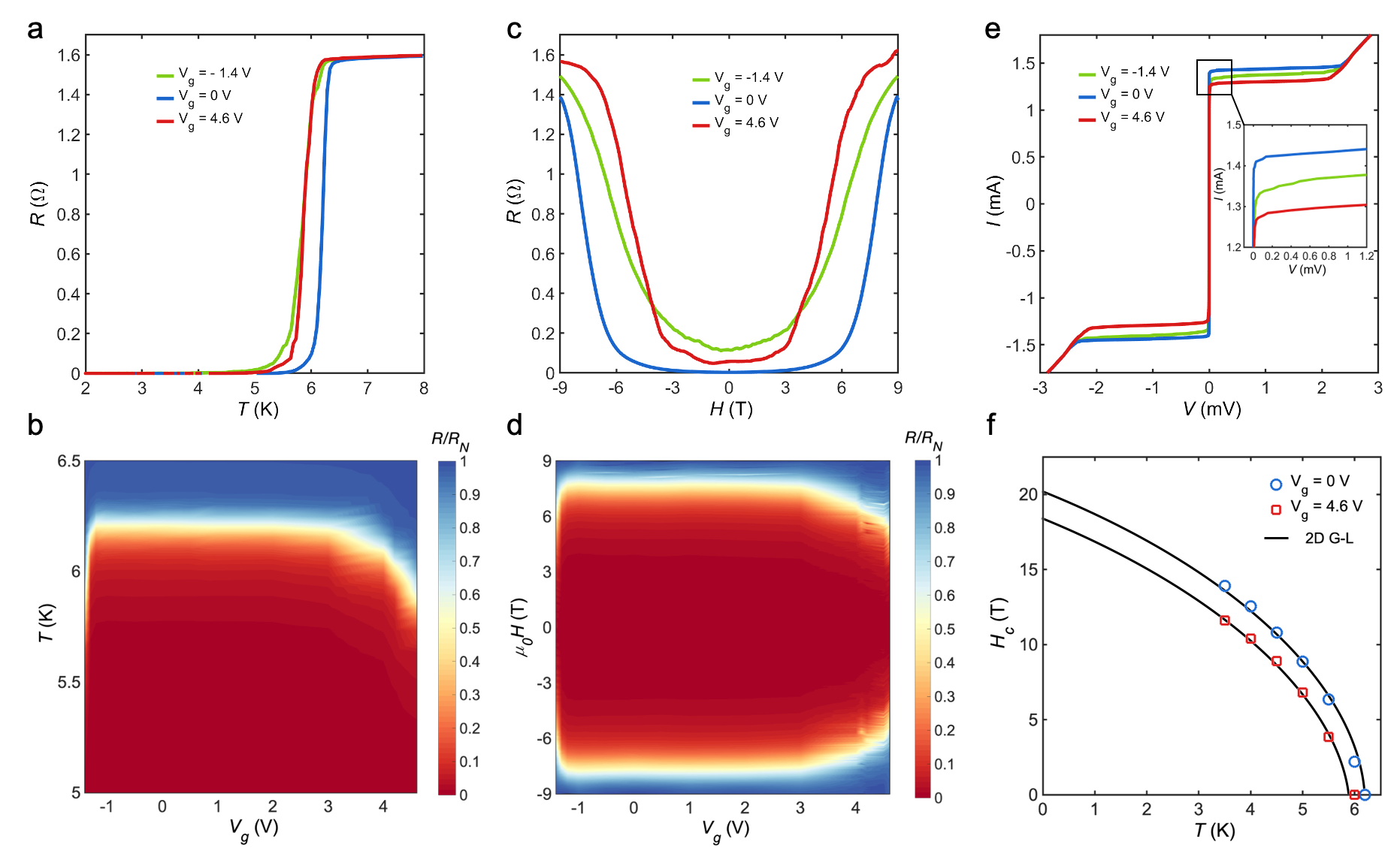}
\caption{{\label{fig2} \textbf{Superconducting characterizations of few-layer NbSe$_2$ under the influence of electrical gating.} \textbf{(a)} The temperature dependence of resistance for NbSe$_2$ under varying $V_{\text{g}}$. \textbf{(b)} Color mapping for the temperature dependence of resistance for NbSe$_2$ under varying gate voltages, where $V_{\text{g}}$ ranges from -1.4 V to 4.6 V. \textbf{(c)} The in-plane magnetic field dependence of resistance for NbSe$_2$ under different gate voltages at a temperature of 5.5 K. \textbf{(d)} Color mapping for the in-plane magnetic field dependence of resistance for NbSe$_2$ under varying gate voltages at a temperature of 5.5 K, where $V_{\text{g}}$ ranges from -1.4 V to 4.6 V. \textbf{(e)} $I$-$V$ curves for NbSe$_2$ under different gate voltages at a temperature of 2 K. The inset provides an enlarged view of the superconducting critical current. \textbf{(f)} The plot shows the in-plane upper critical field as a function of temperature. The black solid line represents the 2D Ginzburg-Landau fitting for the upper critical field $H_{\text{c2}}$.}}
\label{fig:fig2}
\end{figure*}

We first examine the $R$-$T$ curve of the few-layer NbSe$_2$ as shown in Fig. \ref{fig1}(d). The zero-resistance superconducting critical temperature $T_c^{R= 0.1 R_{\text{N}}}$ is approximately 5.9 K and the onset superconducting critical temperature $T_c^{R= 0.9 R_{\text{N}}}$ is approximately 6.25 K. Then, we investigate the electric-field effect in few-layer NbSe$_2$ at a temperature of 5.6 K, where the few-layer NbSe$_2$ enters into a superconducting state. Notably, as the top-gate voltage ($V_{\text{g}}$) is progressively increased from zero, the resistance of NbSe$_2$ also correspondingly increases, gradually enters into the normal state from the superconducting state. Obviously, when the gate current reaches 1 $\mu$A, the resistance of NbSe$_2$ increases rapidly, thereby realizing a switch between the superconducting and normal state. The total current through the NbSe$_2$ of 11 $\mu$A with gate current of 1 $\mu$A is significantly less than the superconducting critical current of 1.4 mA as shown in Fig. \ref{fig2}(e). Meanwhile, as shown in the Supporting Information, changing the test current does not significantly affect the $T_c$ of NbSe$_2$. Therefore, we posit that this modulation effect arises from the injection of quasi-particles induced by the electric field, which will be discussed in detail below. Interestingly, it's noteworthy that a more significant effect of applying negative voltage than that of applying positive was observed in this study, which can be attributed to the inherent ferroelectricity of GaSe flake\cite{BFO1, BFO2, GaSe}. In addition to the work function of graphite and NbSe$_2$ being 4.8 eV\cite{function_G} and 5.5 eV\cite{function_N}, respectively, the polarization of GaSe further changes the Schottky barriers at graphite/GaSe and NbSe$_2$/GaSe interfaces, resulting in a diode effect in this heterostructure. Thus, the leakage current is asymmetric with respect to the gate voltage.

Figure \ref{fig2}(a) presents the temperature-dependent resistance of NbSe$_2$ under three gate voltages. The blue solid line shows a zero-resistance superconducting state of $V_{\text{g}}$= 0 V. When a top-gate voltage of $V_{\text{g}}$= 4.6 V is applied, the $T_c^{R= 0.1 R_{\text{N}}}$ is observed to be suppressed to approximately 5.6 K, as depicted by the red solid line. At a gate voltage of $V_{\text{g}}$= -1.4 V, the critical temperature represented by the green line is also suppressed, similar to the trend observed with the red line. In order to systematically investigate the changes in the critical temperature, Fig. \ref{fig2}(b) presents a color mapping of $R-T$ curves under different $V_{\text{g}}$. After applying the top-gate voltage, the superconducting resistance transition exhibits a systematic change. Indeed, this clearly demonstrates that the superconductivity of NbSe$_2$ can be effectively modulated by the application of a top-gate voltage.

Next, we further investigated the in-plane magnetic field dependence of resistance of NbSe$_2$ under different $V_{\text{g}}$ at $T$ = 5.5 K, as shown in Fig. \ref{fig2}(c) and Fig. \ref{fig2}(d). As anticipated, the in-plane upper critical field of NbSe$_2$ was also reduced under the influence of the top-gate voltage, and the mid-point $H_c$ dropped from 7 T to 4 T at a gate voltage of 4.6 V, which also can be observed in Fig. \ref{fig2}(f). At a gate voltage of 4.6 V, the in-plane upper critical field of the NbSe$_2$ is suppressed throughout the entire superconducting region. Furthermore, the $H_c$-$T$ data for both $V_{\text{g}}$ = 0 V and $V_{\text{g}}$ = 4.6 V were fitted using the 2D Ginzburg-Landau (2D G-L) theory, as represented by the black line. In addition, Fig. \ref{fig2}(e) displays the $I$-$V$ curves of NbSe$_2$ under different $V_{\text{g}}$ at 2 K. Clearly, the superconducting critical current is reduced by 150 $\mu$A and 50 $\mu$A at a gate voltage of 4.6 V and -1.4 V, respectively, which is significantly larger than both the leak current ($I_{\text{leak}}$ = 1 $\mu$A) and the test current ($I_{\text{test}}$ = 10 $\mu$A). 

\begin{figure*}[!htbp]
\centering
\includegraphics[width=1 \linewidth]{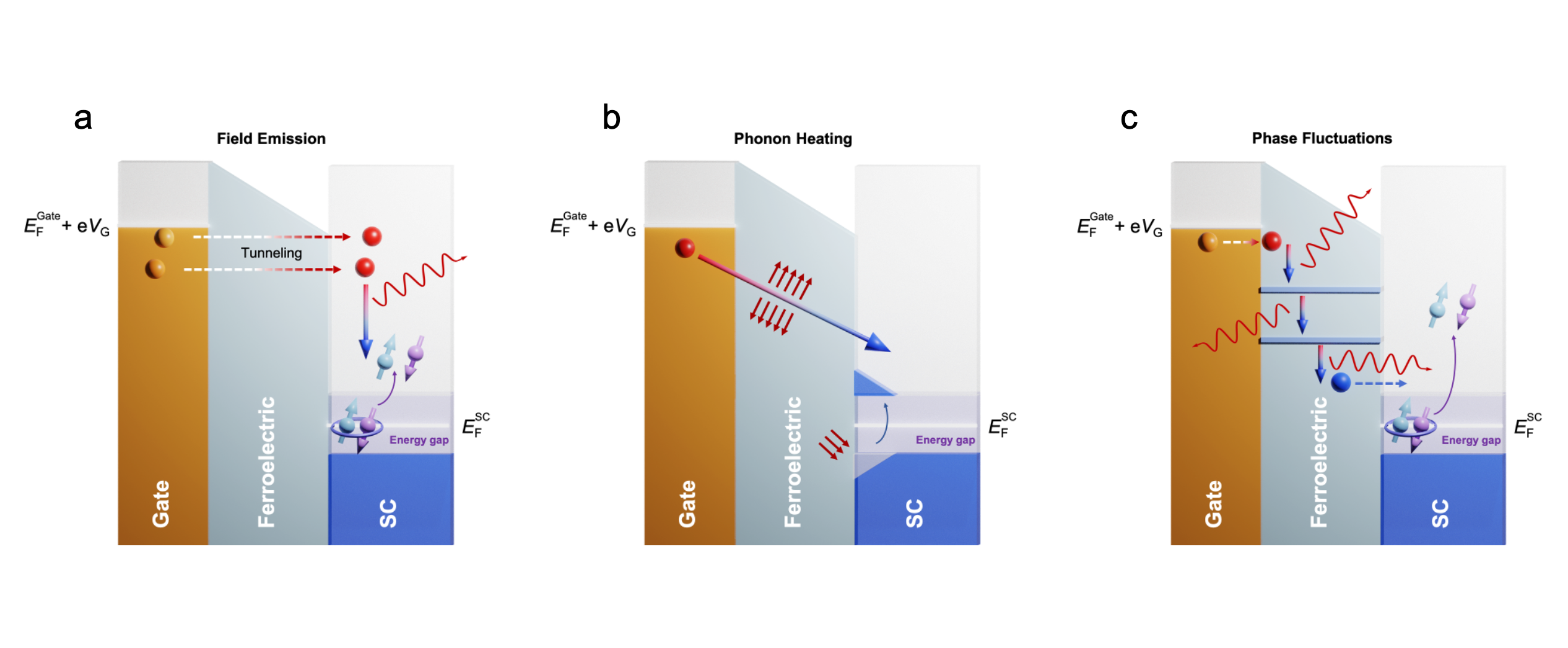}
\caption{{\label{fig3} \textbf{Three mechanisms proposed for the injection of electrons.} \textbf{(a)} Illustration of the `field emission' process of high-energy electrons from a gate electrode into a superconductor, which across the vacuum and relax as phonons, thereby heating the electronic system. \textbf{(b)} Illustration of the `phonon heating' effect which excites electrons in the superconductor. \textbf{(c)} Illustration of the `phase fluctuations' effect, induced by phonons triggered by high-energy electrons traversing from the gate electrode to the superconductor.}} 
\label{fig:fig3}
\end{figure*}

As mentioned above, we posit that this modulation effect arises from the injection of quasiparticles induced by the electric field. In principle, this phenomenon can be explained in three specific mechanisms\cite{Emission1, Emission2, Emission3, Emission4, Phonon1, Phonon2, Phonon3, Phase1, Phase2, arxiv}, as shown in Fig. \ref{fig3}. The first scenario is the `field emission' or `direct tunneling' effect of high-energy electrons from the gate electrode into the superconductor across the dielectric layer\cite{Emission1, Emission2, Emission3, Emission4}, as shown in Fig. \ref{fig3}(a). The hot electrons are injected into the superconductor then relax as phonons or quasiparticles within it, heating the electronic systems and thereby affecting the superconducting properties. In the second scenario (Fig. \ref{fig3}(b)), the electrons in the superconductor are also heated. However, this heating effect differs from vacuum tunneling and instead refers to the phonon-mediated heating by charge carriers leaking through the dielectric layer, often referred to as `phonon heating'\cite{Phonon1, Phonon2, Phonon3}. In the third scenario (Fig. \ref{fig3}(c)), `phase fluctuations' effect describes a non-equilibrium state in the superconductor caused by phonons and/or high-energy electrons injected into the superconductor, but without considerable heating of the electronic system\cite{Phonon3, Phase1, Phase2}. However, the differences between scenario 1, 2, and 3 are very subtle and difficult to distinguish experimentally. Consequently, they are often collectively referred to as the high-energy electrons injection effect. In our case, the thickness of GaSe layer is much larger than the distance that tunneling could happen, making scenario 2 or 3 more likely. As mentioned previously, while this phenomenon has been observed in various superconducting thin films and nanowires, this is the first instance of its observation in 2D superconductors.

\begin{figure*}[!htbp]
\centering
\includegraphics[width=0.8\linewidth]{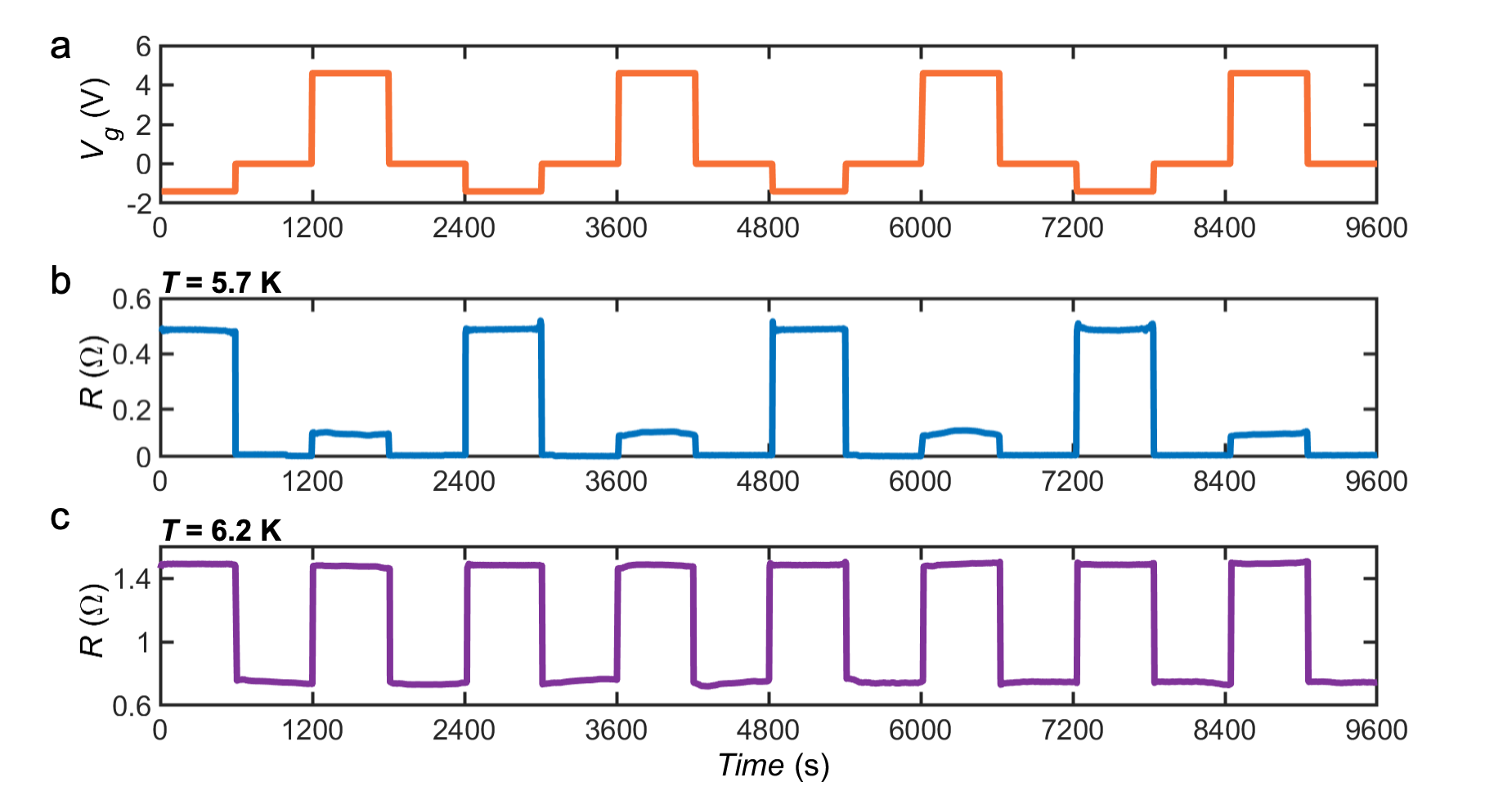}
\caption{{\label{fig4} \textbf{Electrical switching between the superconducting and normal states in a NbSe$_2$/GaSe heterostructure.} \textbf{(a)} A sequence diagram of applied top-gate voltages of $V_{\text{g}}$ = -1.4 V, 0 V, 4.6 V, and 0 V, for switching between different states in the NbSe$_2$/GaSe heterostructure. \textbf{(b)} Illustration of reversible switching for resistance over multiple cycles at a temperature of 5.7 K. The switching between the metallic and superconducting states has been realized.\textbf{(c)} Illustration of reversible switching for resistance over multiple cycles at $T$ = 6.2 K. The switching between different metallic states has been realized.}}
\label{fig:fig4}
\end{figure*}

Therefore, based on the above measurements, it can be concluded that whether it is the superconducting critical temperature, the superconducting critical magnetic field, or the superconducting critical current, all can be effectively modulated by the gate-controlled injection of quasi-particles. Having established the manipulation of superconducting properties and understood the mechanism, we now exploit this unique property to achieve reversible switching between these distinct electronic states, which could have significant implications for the design of future superconducting circuits and devices. In Fig. \ref{fig4}, we initially preconditioned the system with a top-gate voltage of $V_{\text{g}}$= -1.4 V before setting $V_{\text{g}}$= 0 V. Remarkably, as shown in Fig. \ref{fig4}(b), we demonstrated the switch for the resistance from a superconducting state to a stable metallic state by applying a voltage of $V_{\text{g}}$= -1.4 V. After the system was stable in the metallic state, we then removed the applied electric field. Notably, after the removal of the electric field, the system returned to a superconducting state. Subsequently, we applied a second electric field with a voltage of  $V_{\text{g}}$= 4.6 V, driving the device into another metallic state. To underscore the stability of this procedure, we performed a sequence of alternating gate voltages: $V_{\text{g}}$= -1.4 V, 0 V, 4.6 V, and 0 V. This allowed us to switch the system between the superconducting and metallic states repeatedly at 5.7 K, as shown in Fig. \ref{fig4}(b). Similarity, we repeated this operation on the voltage at higher temperature (as shown in Fig. \ref{fig4}(c) at $T$ = 6.2 K), it can be observed that the device exhibits stable and efficient switching between different metallic states. This demonstrates the robust bistable behavior present in this system and illustrates its potential as a van der Waals platform for superconducting switches.

\begin{figure*}[!htbp]
\centering
\includegraphics[width=1 \linewidth]{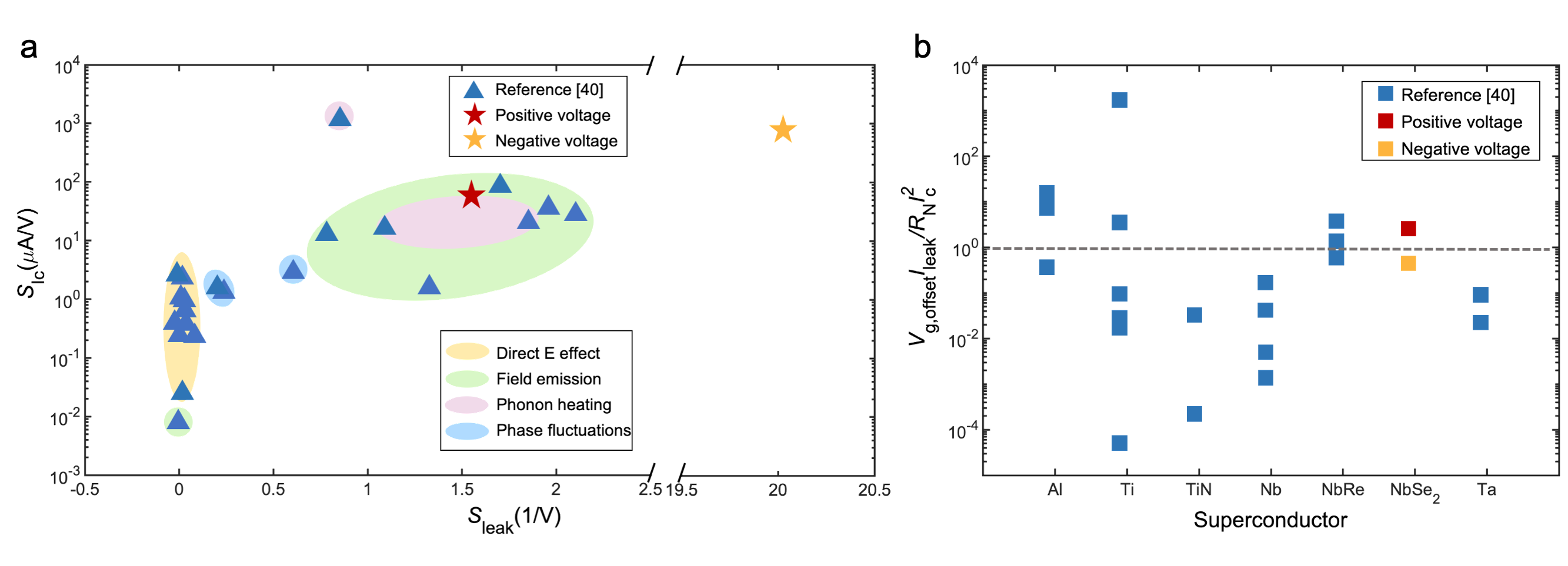}
\caption{{\label{fig5} \textbf{Compare the dissipated power of GaSe/NbSe$_2$ device with different studies.} \textbf{(a)} $S_{\text{Ic}}$ as a function of $S_{\text{leak}}$ is shown for different studies, represented by colored bubbles, each corresponding to a different scenario of the high-energy electrons injection effect. \textbf{(b)} Power dissipation ratio in different studies. All blue marks in \textbf{(a)} and \textbf{(b)} are copied from the reference\cite{arxiv}.}} 
\label{fig:fig5}
\end{figure*}

In order to evaluate the performance of the devices and compare with different studies, we also calculate the parameter $S_{\text{leak}}$ = (Log ($I_{\text{leak,offset}}$/$I_{\text{leak,onset}}$))/($V_{\text{g,offset}}$-$V_{\text{g,onset}}$) from the $I_{\text{leak}}$-$V_{\text{g}}$ curve and the $S_{\text{Ic}}$ = ($I_{\text{c,onset}}$-$I_{\text{c,offset}}$)/($V_{\text{g,offset}}$-$V_{\text{g,onset}}$) from the $I_{\text{c}}$-$V_{\text{g}}$ curve\cite{arxiv}, respectively. These two parameters represent the steepness of $I_{\text{leak}}$ and $I_{\text{c}}$. In the context of positive voltage, it can be obtained that the $S_{\text{leak}}$ is around 1.53, indicating that the $I_{\text{leak}}$ trends follows an exponential increase or a power low with a large exponent ($I_{\text{leak}}$ $>$ 1), thus further eliminates the direct electric field effect. Meanwhile, the $S_{\text{Ic}}$ is around 57.7, signifying a rapid decline in $I_{\text{c}}$ with increasing $V_{\text{g}}$. Using the two parameters $S_{\text{leak}}$ and $S_{\text{Ic}}$, we can compare our results with other studies as shown in Fig. \ref{fig5}(a) (marked by the red star), and roughly estimate that the phonon heating effect plays an important role in the NbSe$_2$/GaSe vdW device\cite{arxiv}. Furthermore, to quantify this contribution, we have also calculated the power dissipation ratio $P_{\text{g,offset}}$/$P_{\text{N}}$ as a function of the superconducting NbSe$_2$, as shown in Fig. \ref{fig5}(b). $P_{\text{g,offset}}$/$P_{\text{N}}$ is defined as $V_{\text{g,offset}}$$I_{\text{leak}}$/$R_{\text{N}}$$I_{\text{c}}^2$, with a calculated value of approximately 1.46, which is similar to the results observed in gate-controlled titanium nanobridge supercurrent transistor\cite{Nanobridge}.

Similarly, we also calculate the parameters under negative voltage conditions. Notably, the values of $S_{\text{leak}}$ and $S_{\text{Ic}}$ are 20 and 750, respectively, as marked by the orange star in Fig. \ref{fig5}(a). The two parameters remarkably exceed those reported in previous works, which may be due to the diode effect of the heterostructure. Additionally, our calculations of the power dissipation ratio indicate that it is significantly reduced under the negative gate voltage, approximately to 0.45 as shown in Fig. \ref{fig5}(b). This low power dissipation ratio is comparable to that reported in Nb Dayem bridge\cite{Nbbridge}.

Therefore, we have broken the symmetric suppressing effect of $I_{\text{c}}$ on $V_{\text{g}}$ reported in previous works\cite{arxiv}, and we attribute this phenomenon to the ferroelectric polarization from the 2D GaSe flake. In the background of ferroelectric polarization field, the phonon heating and phase fluctuations effect play an important role under positive voltage and become more efficient at the negative side. But fundamentally, they both arise from the enhancement of non-equilibrium phase fluctuations due to the injection of high-energy electrons via electric field. It is interesting to further explore the effect of polarization reverse by a high gate voltage pulse, potentially constructing a superconducting switch with memory effect. We anticipate that integrating novel quantum 2D materials with 2D superconductors will not only facilitate the development of devices with reduced power consumption but also deepen our comprehension of the quasi-particle injection effect.


\section{CONCLUSION}
In summary, we successfully fabricated a top-gate tuning device consisting of a superconducting few-layer NbSe$_2$ and a dielectric GaSe flake. By applying an electric field to inject high-energy electrons, we effectively modulated the superconductivity of few-layer NbSe$_2$. We observed a steep gate dependence by realizing a ferroelectric diode at the gate dielectric. Furthermore, we utilized this singular device to achieve a superconducting switch capable of multiple cycles at a temperature of 5.7 K. This developed gating method employing GaSe holds great promise for application in the study of other vdW layered superconductors, thereby unlocking the potential of nanoscale superconducting circuits.

\section{METHODS}
\textbf{Device Fabrication}. Bulk GaSe was purchased from Nanjing MKNANO Tech. Co., Ltd. (www.mukenano.com). The devices were fabricated by a dry-transfer technique under vacuum condition. Typically, thin graphite flake, GaSe flake, and few-layer NbSe$_2$ were mechanically exfoliated from high-quality single crystals by polydimethylsiloxane (PDMS), respectively. The flakes were sequentially transferred and precisely aligned onto the as-prepared substrates, as depicted in an optical image in Fig. \ref{fig1}(c), where the blue, red  and white solid lines represent the few-layer NbSe$_2$, thin graphite flake, and GaSe flake, respectively. 

The as-prepared electrodes were fabricated by a standard photolithography method. Typically, the pattern was written onto the SiO$_2$/Si substrates by laser direct-write lithography system (MicroWriter ML3) with 365 nm long-life semiconductor light source. Subsequently, the electrode patterns were etched to a depth of 25 nm with CHF$_3$ plasma generated from a reaction ion etching (RIE) system. Finally, Ti/Au (5nm/20nm) was deposited via an electron beam deposition (EBD) system. As shown in Fig. \ref{fig1}(b), the electrodes were embedded into the SiO$_2$/Si to form a flat substrate, thereby minimizing the contact resistance between the few-layer NbSe$_2$ and the electrodes.\\
\textbf{Measurements}. The device was characterized using a Physical Property Measurement System (PPMS, Quantum Design). Superconducting signals were measured using a Keithley 2400 and 2182A as the current source and voltage meter, respectively. A low current of 10 $\mu$A was applied to prevent local heating and current-driven effects. 

\section{ASSOCIATED CONTENT}
\textbf{Supporting Information}. \\
$R$-$T$ curves of device with different test currents. 

\section{AUTHOR INFORMATION}
\textbf{Corresponding Author}. \\
Jun Li -- School of Physical Science and Technology, ShanghaiTech University, Shanghai 201210, China. Email: lijun3@shanghaitech.edu.cn\\
Yueshen Wu -- School of Physical Science and Technology, ShanghaiTech University, Shanghai 201210, China. Email: wuysh@shanghaitech.edu.cn\\
Lan Chen -- Institute of Physics, Chinese Academy of Sciences, Beijing, 100190, China. Email: lchen@iphy.ac.cn\\

\noindent{\textbf{Author  Contributions}}\\
All authors discussed the results and commented on the manuscript.

\noindent {\bf Notes}\\
\noindent The authors declare no competing financial interests.

\bigskip
\section{ACKNOWLEDGEMENT}

This research was supported in part by the Ministry of Science and Technology (MOST) of China (No. 2022YFA1603903), the National Natural Science Foundation of China (T2325028, 12134019, 12104302, 12104303), and the Science and Technology Commission of Shanghai Municipality. The authors also are thankful for the support from the Soft Matter Nanofab (SMN180827), SPST, ShanghaiTech University.

%



\providecommand{\latin}[1]{#1}
\makeatletter
\providecommand{\doi}
  {\begingroup\let\do\@makeother\dospecials
  \catcode`\{=1 \catcode`\}=2 \doi@aux}
\providecommand{\doi@aux}[1]{\endgroup\texttt{#1}}
\makeatother
\providecommand*\mcitethebibliography{\thebibliography}
\csname @ifundefined\endcsname{endmcitethebibliography}
  {\let\endmcitethebibliography\endthebibliography}{}

\clearpage

\end{document}